\documentclass[12pt]{article}
\usepackage[export]{adjustbox}
\usepackage{amsmath,latexsym}
\usepackage{physics}
\usepackage{verbatim}
\usepackage{hyperref}
\usepackage{graphicx}
\usepackage{authblk}
\usepackage{amssymb}
\usepackage{cite}
\newcommand{\R}{\mathbb{R}}

\usepackage{a4wide}
\usepackage{xcolor}
\usepackage{caption,subcaption}
\title{A Note on Convex Realization of Halohedron\\\vspace*{1cm}}
\date{}
\author{Saroj Chhatoi }
\affil{Institute of Mathematical Sciences, Taramani, Chennai 600 113, India \vspace*{-.4cm}}\affil{Homi Bhabha National Institute, Anushakti Nagar, Mumbai 400085, India}
\affil{ E-mail: \href{mailto:sarojpc@imsc.res.in}{sarojpc@imsc.res.in} }

\usepackage{imakeidx}
\makeindex

\begin{document}

\maketitle

\begin{abstract}
	In the recent works \cite{Salvatori_2018},\cite{salvatori20181loop}, Halohedron emerged as amplituhedron for 1-loop planar diagrams in bi-adjoint massless $\phi^3$ theory. Halohedron is a specific case of graph cubeahedron where the considered graph is a cycle-graph. In \cite{padrol2019associahedra},\cite{manneville2015compatibility}, the authors provide construction of any graph cubeahedron and we use this construction to find the polytopal realization of Halohedron. We show that the Halohedron we obtain is equivalent to the proposed realization of Halohedron in `Big Kinematic Space'\cite{salvatori20181loop}. 
	
\end{abstract}
\newpage    
\tableofcontents

\section{Introduction}
The recent developments in `\textit{Amplituhedron program}' have established connections between positive geometries and S-matrix for various class of theories. \cite{Arkani_Hamed_2018} Nima et. al. show that the canonical form of a polytope known as Associahedron sitting inside the kinematic space determines the amplitude for planar tree-level diagrams in massless bi-adjoint $\phi^3$ theory. The polytopal realization of Associahedron is well studied in mathematical literature and in recent seminal work\cite{padrol2019associahedra}, it was shown that the realization provided in \cite{Arkani_Hamed_2018} is the same as the `simplicial type cone' based realization of Associahedron. 

An attempt to extend the `\textit{Amplituhedron program}' to the analysis of planar 1-loop amplitude for bi-adjoint $\phi^3$ theory  \cite{Salvatori_2018},\cite{salvatori20181loop} led to   Halohedron as the corresponding polytope. Halohedron is a specific case of a general class of convex polytopes known as graph cubeahedron. In \cite{padrol2019associahedra}, the authors present a polytopal realization of graph-cubeahedron using objects known as $g$-vector fans. In this paper, we use this construction to find the polytopal realization of Halohedron.
Similar to the case of massless bi-adjoint $\phi^3$ theory, a natural question would be to ask if the realization of Halohedron proposed in \cite{salvatori20181loop} matches with that obtained from the $g$-vector fans of Halohedron. 

This paper aims to derive the facet defining inequalities of Halohedron from the theory of $g$-vector fans. The facet defining inequalities will be linear functions which will serve to embed the Halohedron in an ambient space $\R^n$. We find that  \textit{g}-vector fan-based realization of Halohedron is the same as the realization provided in \cite{salvatori20181loop}. On the way of achieving this goal, we briefly digress into the theory of \textit{g}-vector fans needed to derive the inequalities. The linear inequalities proposed in \cite{salvatori20181loop},    come with constants $\epsilon$'s which should satisfy some constraints if the proposed inequalities are to be consistent with $g$-vector fan \cite{padrol2019associahedra}  based realization. We provide the proof that these $\epsilon$'s do satisfy the same constraints and hence this will complete the convex realization of Halohedron. 

The paper is organized as follows:
We start by summarizing Salvatori's construction of Halohedron in space of abstract variables. These abstract variables in one way correspond to 1-loop propagators where we have not yet imposed momentum conservation conditions.   In section (\ref{sec3})  we introduce graph associahedra and graph cubeahedron. In doing so we introduce tubings and use these tubings to iteratively truncate a $n$-cube to obtain a graph cubeahedron. In section (\ref{sec4}), without delving into the details we introduce the language of $g$-vector fans necessary enough to motivate the convex realization of Halohedron. In section (\ref{sect5}) we use the $g$-vector fans of graph cubeahedron to derive the facet defining inequalities of Halohedron. We give illustrative proofs for $n=2$, $n=3$ and arbitrary $n$ case. In section (\ref{sec6}) we show that the constants $\epsilon$ appearing in the facet defining inequalities satisfy constraints on the height vectors in the \textit{g}-vector language. This completes the convex realization of Halohedra.

\section{Facet Defining Inequalities of Halohedron}
In \cite{salvatori20181loop},\cite{Salvatori_2018} the authors proposed Halohedron sitting inside a kinematic space as the convex polytope encoding the 1-loop amplitude for planar $\phi^3$ theory. 
The Halohedron is first embedded in a 'Big Kinematic Space' where the variables $X_I$ corresponding to the facets are abstract variables where the momentum conservation is yet not imposed. 


The Halohedron will be embedded in $\R^n$ by imposing positivity on a set of linear functions `$X_I$'. These linear functions are in $1-1$ correspondence to the facets of the Halohedron and thus to the propagators of 1-loop planar diagrams\cite{jagdale},\cite{salvatori20181loop}. By imposing positivity conditions on these variables we ensure that Halohedron defines a region where all the propagators are positive. 

Let $X_1,X_2,..,X_n \in \mathbb{R}^n$  be the set of independent variables, such that $(X_1,X_2, \hdots, X_n)$ be any generic point and define $X_{(i,i+1)}$ related to $X_i$ as
\begin{equation}
X_{(i,i+1)}:= \epsilon^1_{(i,i+1)} - X_{i}\label{eqn4:Xi}
\end{equation}
The variables $X_i$ and $X_{(i,i+1)}$ are associated with the opposite facets of $n$-cube which is to be truncated.

We first truncate the vertex of $n$-cube corresponding to intersection of all the $X_{(i,i+1)}$ facets $i.e.$ $\cap_{i=1,2 .. ,n}X_{i,i+1}$ by imposing positivity conditions on $X_0$ which is defined as,
\begin{equation}
X_0:=\sum_{i=0}^n X_{(i,i+1)} - \epsilon^n_0
\end{equation}
Then we further truncate the polytope at intersection of the $X_{(i,i+1)}$ facets in order of increasing dimensions, by imposing positivity on the following variables
\begin{equation}
X_I:=\sum_{j\in I'}X_{(j,j+1)}-\epsilon_I^{|I'|}
\end{equation}
where $I$ is a subset of cyclically consecutive indices, $I'$ is the same subset with the last index removed and $|I'|$  denotes the cardinality of $I'$. For example, in the case of $n=4$, $X_{(1,2,3)}:=X_{(1,2)}+X_{(2,3)}-\epsilon^2_{(1,2,3)}$ is one such variable for which $I=\{1,2,3\}$ and $I'=\{1,2\}$. Here $X_{(1,2,3)}$ truncates the polytope at the facet formed by the intersection of $X_{(1,2)}$ and $X_{(2,3)}$. \textbf{Note}: All of these constants $\epsilon$ are positive constants but cannot be chosen  arbitrarily  in order avoid deep cuts while truncating the facets. We will see in later sections that these constants $\epsilon$ indeed satisfy some constraints.

This truncated $n$-cube is the desired Halohedron $H_n$.
The vertices on the Halohedron separated by an edge corresponds to Feynman diagrams related by mutation\cite{salvatori20181loop}. So one fixes one of the diagrams as the reference diagram and the measure of all other vertices will be related by mutation up to an overall sign. 
The canonical form of Halohedron $H_n$ then can be written as a sum over all 1-loop planar diagrams 
\begin{equation}
\Omega_{H_n}=\pm d\mu_{g*} \left\{\sum_g \frac{1}{\prod_{I\in g} X_I} \right \}
\end{equation}
where $g^*$ is the fixed reference diagram, $d\mu_g=\bigwedge_{I\in g^*} dX_I$ and in $\prod_{I\in g}$ $I$ runs over all the propagators of diagram $g$ .  
To obtain the 1-loop amplitude from this canonical form we strip out the measure $d\mu_g^*$. Though in this case, the remaining factor will contain unphysical contributions from diagrams containing external bubble and tadpole, we can kill such contributions by sending the corresponding unphysical variables to infinity. Then we impose momentum conservation on the variables $X_I$  to obtain the 1-loop planar amplitude for massless $\phi^3$ theory.

\section{Graph Associahedra} \label{sec3}

In this section we summarise the construction of graph associahedra provided in \cite{devadoss2006realization},\cite{devadoss2010deformations}.
Let \textbf{G} be a connected graph with the corresponding vertex set $V$. A $tube$ is a set of nodes of graph whose induced graph is a connected subgraph of \textbf{G}.

The tubes which are neither empty nor maximal (containing all the nodes) are called $proper$. In what follows tubes are assumed to be proper. Two tubes $u_1$ and $u_2$ interact on the graph as follows: 
\begin{itemize}
	\item Tubes are \textit{nested} if $u_1 \subset u_2$.
	\item Tubes \textit{intersect} if $u_1 \cap u_2 \neq \emptyset$ and $u_1\not\subset u_2 $ and $u_2\not\subset u_1$ 
	\item Tubes are \textit{adjacent} if $u_1\cap u_2=\emptyset$ and $u_1 \cup u_2$ is a tube in \textbf{G}.
\end{itemize}
A pair of tubes are said to be \textit{compatible} if they do not intersect and are not adjacent. A tubing $U$ of \textbf{G} is a set of tubes of \textbf{G} such that every pair of tubes in $U$ is compatible.

\begin{figure}[h] 
	\centering 
	\includegraphics[scale=0.5,width=\linewidth]{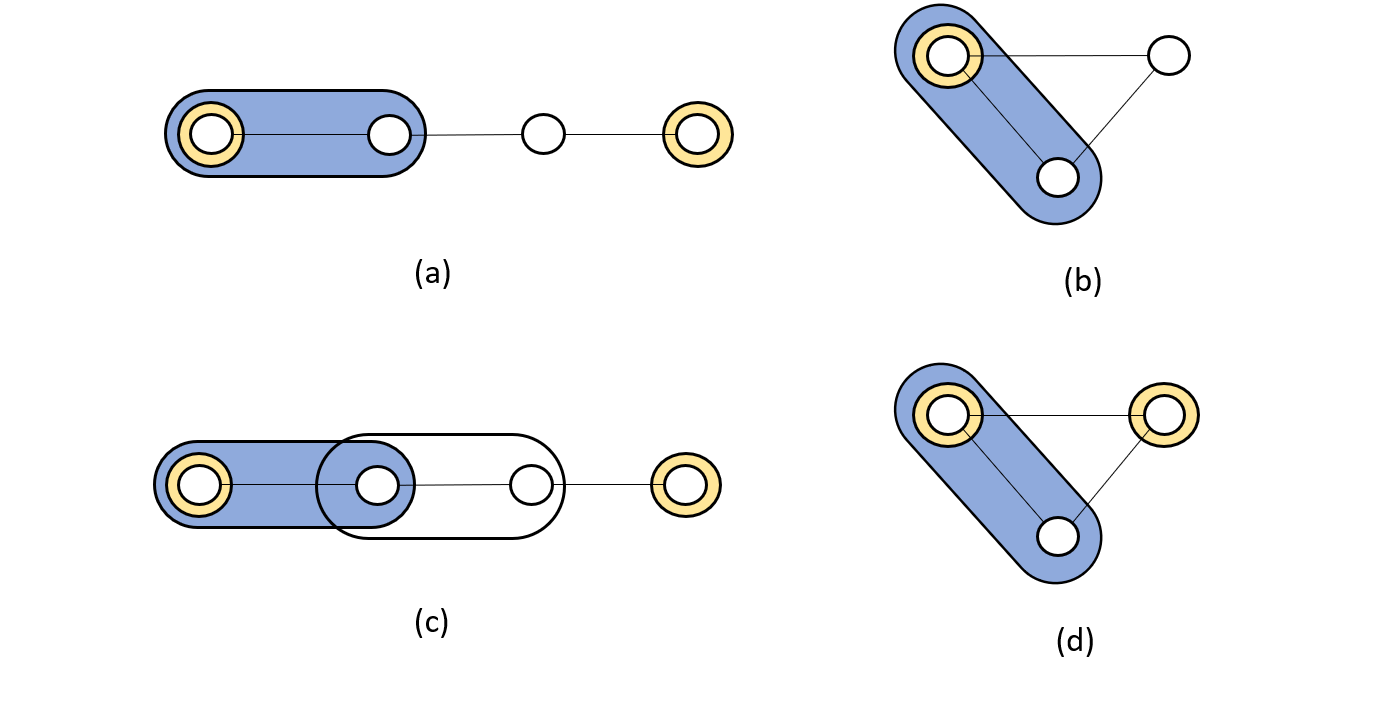}
	\caption{The example of tubes. Figure ($a$) and ($b$) are examples of compatible tubes whereas figure ($c$) and ($d$) are examples of set of  incompatible tubes} \label{fig1:comp}
\end{figure}

In Figure (\ref{fig1:comp}), (a) and (b) show two examples of compatible tubes but (c) and (d) are examples of incompatible tubes as there are adjacent and intersecting tubes in them.
Using tubings we will construct graph associahedra $\mathcal{K}G$ as follows.
\begin{itemize}
	\item Let $\Delta_n$ be a $(n-1)$ simplex such that each co-dimension one facet of $\Delta_n$ labels a particular node of graph.  
	\item Starting with the corner vertex iteratively truncate the simplex such that new facets are associated with tubings of higher dimension $i.e.$ tubings obtained by adding more tubes. 
\end{itemize}

Below we provide an illustrative example for the case of a cycle graph consisting of 3 nodes.
\begin{figure}[hbt!]
	\includegraphics[scale=0.45,width=\linewidth]{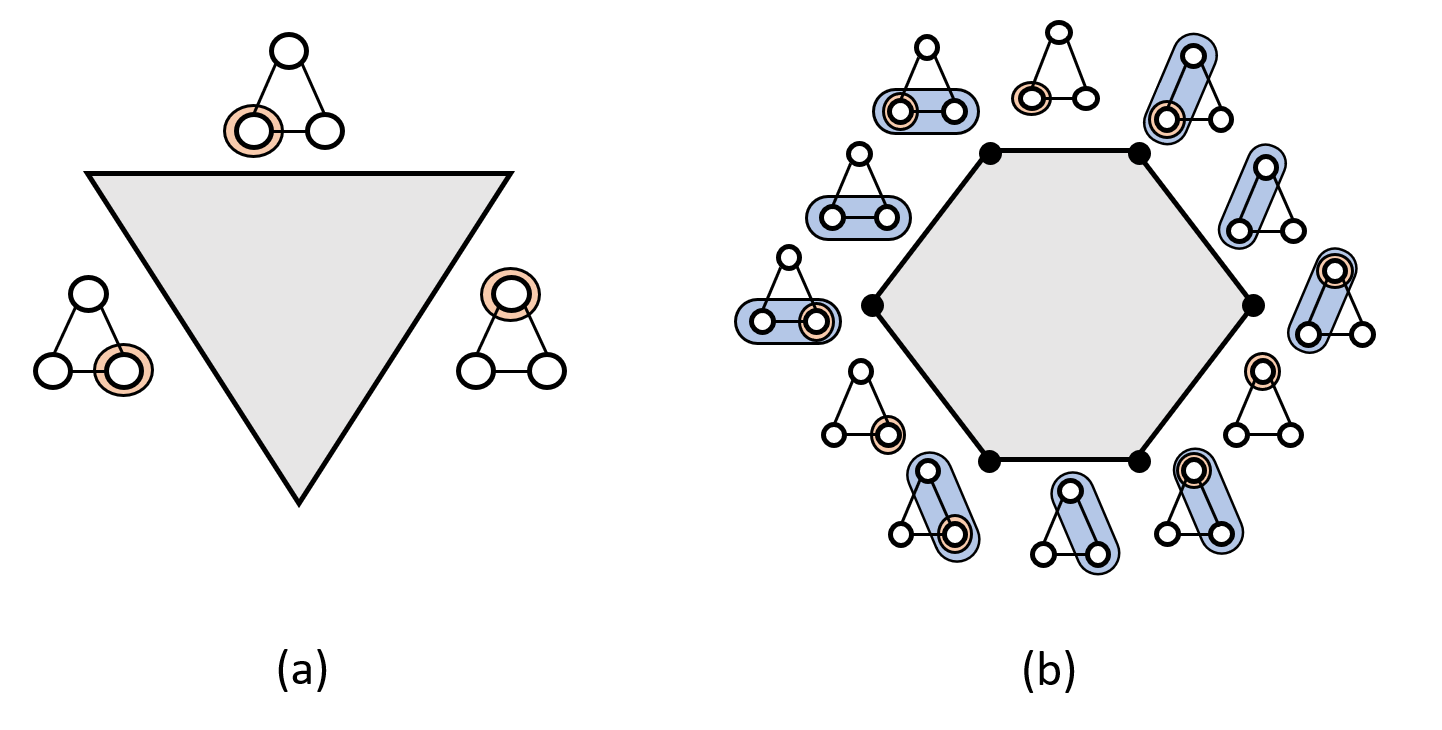}
	\caption{Truncation of 2-simplex to obtain $W_2$} \label{fig2:comp}
\end{figure}
In figure (\ref{fig2:comp}) (a), we start with a 2-simplex where each edge represents a particular node of the graph. Then we truncate every vertex of the 2-simplex such that each new facet corresponds to a higher dimensional tubing (tubing with 2 nodes). In fig 2 (b), we obtain cyclohedron $W_2$ where, the vertices of polyhedron represent highest dimensional tubings (tubings consisting of proper tubes) possible for the cycle graph. For a \textit{cycle}-graph with arbitrary $n$ nodes $\mathcal{K}G$ becomes a cyclohedron, $W_n$. Similarly for a \textit{path}-graph with $n$ nodes one obtains associahedron $K_{n+1}$. In general for a graph with $n$ nodes one obtains a graph associahedron which is a $simple$ $convex$ $polytope$ of $n-1$ dimensions.
For further details of different graph associahedra for example cyclohedron, associahedron we refer the reader to  \cite{devadoss2006realization}.   
\\

To motivate the construction of Halohedron which is the object of interest of this paper we define \textit{design tubing} and a new class of polytope called \textit{graph cubeahedron} which is obtained by iterated truncation of a $n$-cube such that its face poset is isomorphic to \textit{design tubings}. \\ \\
Design tubes are composed of two kinds of tubes:
\begin{itemize}
	\item The round tubes: these are usual tubes of \textbf{G} introduced earlier but not necessarily proper.
	\item The square tubes: these are just single nodes of \textbf{G}.
\end{itemize}
Following the similar idea of compatibility of tubes discussed above, we define the compatibility of design tubes.
A pair of design tubes are compatible if:
\begin{itemize}
	\item Both are round tubes, are not adjacent and do not intersect
	
	\item Or at least one of them is a square tube and they are not nested.
\end{itemize}
A $design$ $tubing$ $U$ of $\mathbf{G}$ is a collection of design tubes of $\mathbf{G}$ such that every pair of tubes in $U$ is compatible. 
\begin{figure}[hbt!]
	\includegraphics[scale=0.45,width=\linewidth]{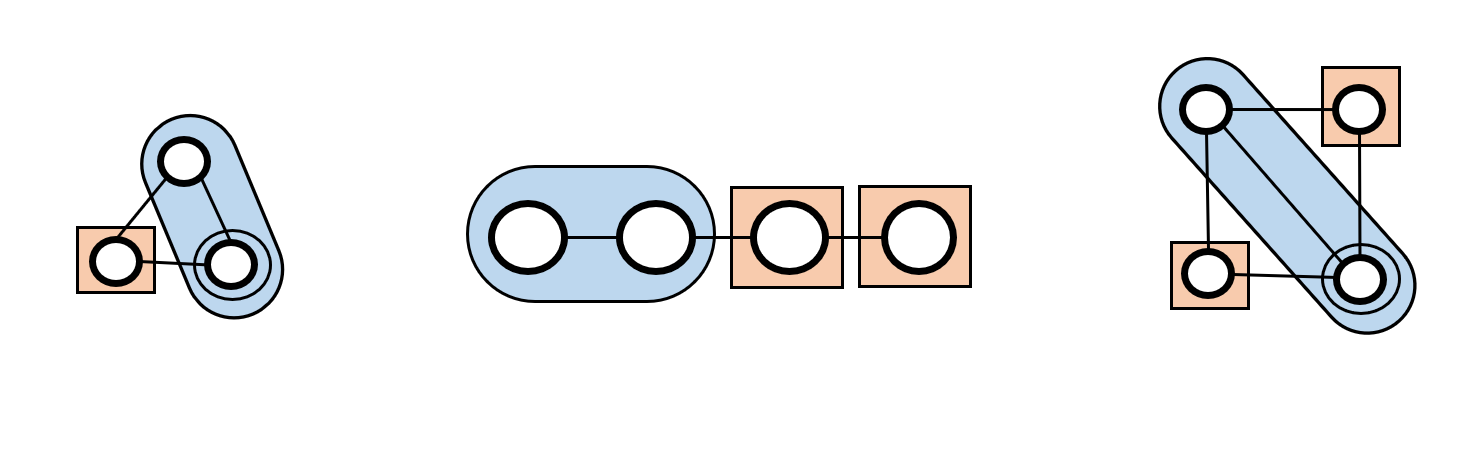}
	\caption{Examples of compatible design tubes} 
\end{figure}
\\
In order to construct Halohedron we will use these design tubings of cycle graph to iteratively truncate $n$-cube.
Below we sketch the algorithm:
\begin{itemize}
	\item Take a $n-$cube with each pair of its opposite facets represents one node of the graph. One of the facets of each such pair represents the node with a square tube and the other corresponds to the node with a round tube.  
	
	\item Now starting with the corner vertex, adjacent to facets corresponding to only round tubes, truncate the cube such that each new facet maps to higher dimensional tubings with round tubes.
\end{itemize} 

\begin{figure}[h]
	\includegraphics[scale=0.45,width=\linewidth]{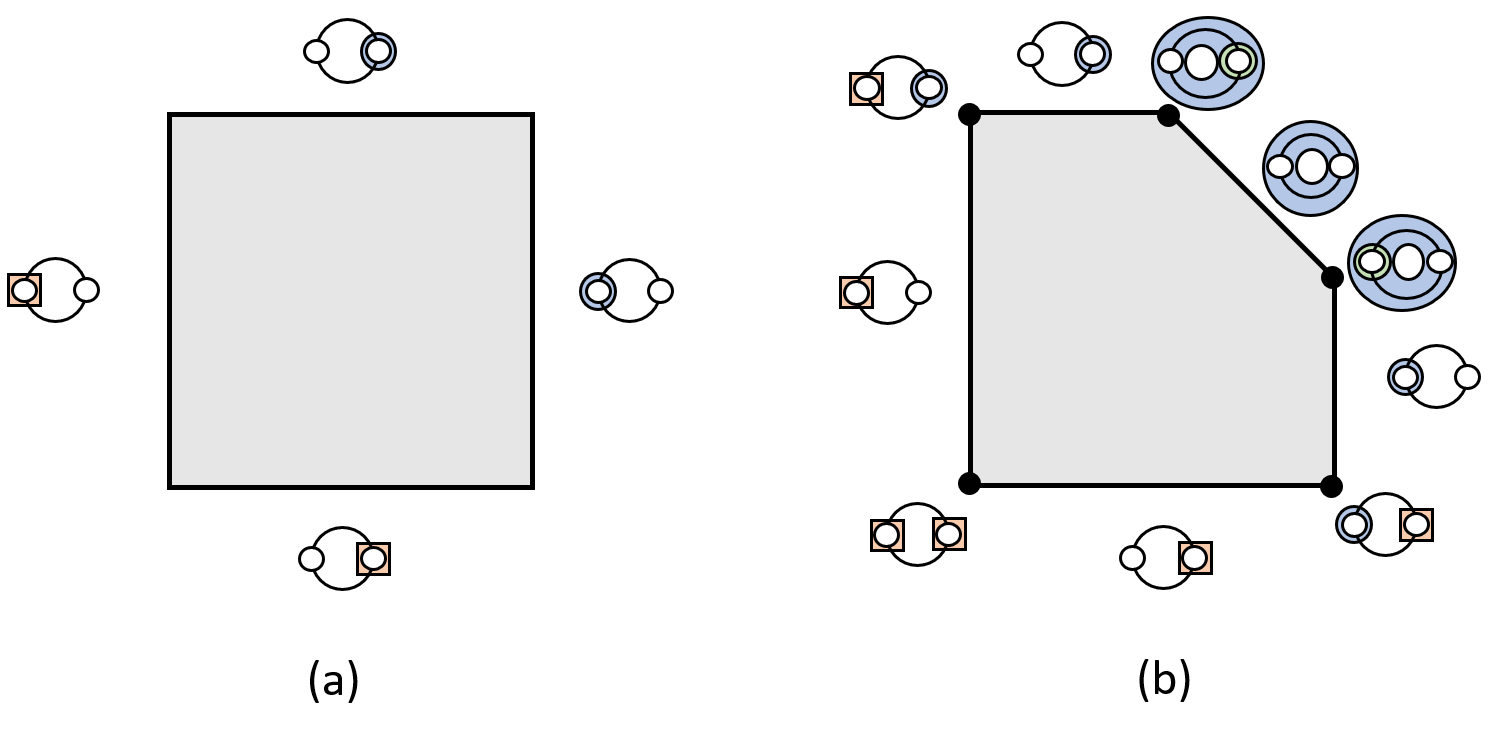}
	\caption{Iterative truncation of 2-cube to obtain graph cubeahedron for cycle graph with 2 nodes ($H_2$)} 
\end{figure}
We give an illustrative example of a graph cubeahedron obtained by the iterated truncation of a 2-cube. Start with a 2-cube where the pair of opposite facets of the 2-cube corresponds to a particular node of the graph. Then truncate the corner vertex and the new facet formed is labeled by the round tube enclosing the whole graph. In figure 4 (b) the vertex of $H_2$ are labeled by tubings which enclose the entire graph and it need not be a proper tube which contrasts the case of graph associahedron where the tubes were proper.    

\begin{figure}[!htb]
	\centering
	\includegraphics[scale=.4]{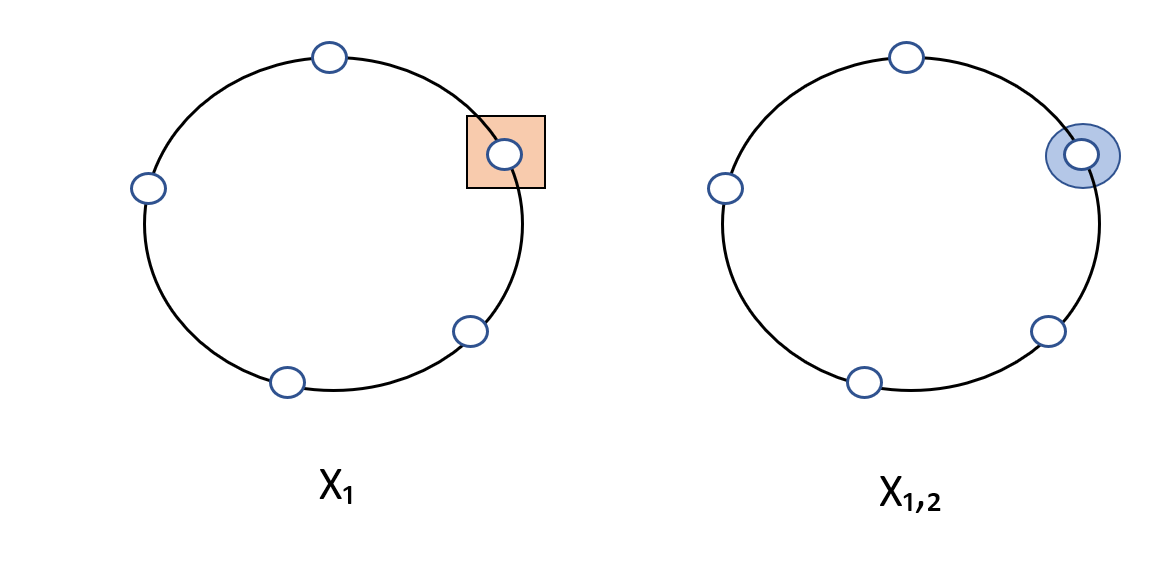}
	\caption{An example of variable $X_i$ and $X_{(i,i+1)}$ for the case of n=5}
\end{figure}

We note that the variables $X_I$'s used in Salvatori's construction  are in $1-1$ correspondence with the \textit{design tubings} of $cycle$-graph. The variables $X_i$ correspond to tubings with a single square tube on one of the nodes and $X_{(i,i+1)}$ corresponds to tubing with a round tube on the same node as shown in figure 5. Also, note that $X_I$ can be related to design tubings with round tubes consisting of a set of $|I|-1$ cyclically consecutive nodes of the graph.

\begin{figure}[!htb]
	\centering
	\includegraphics[scale=.4]{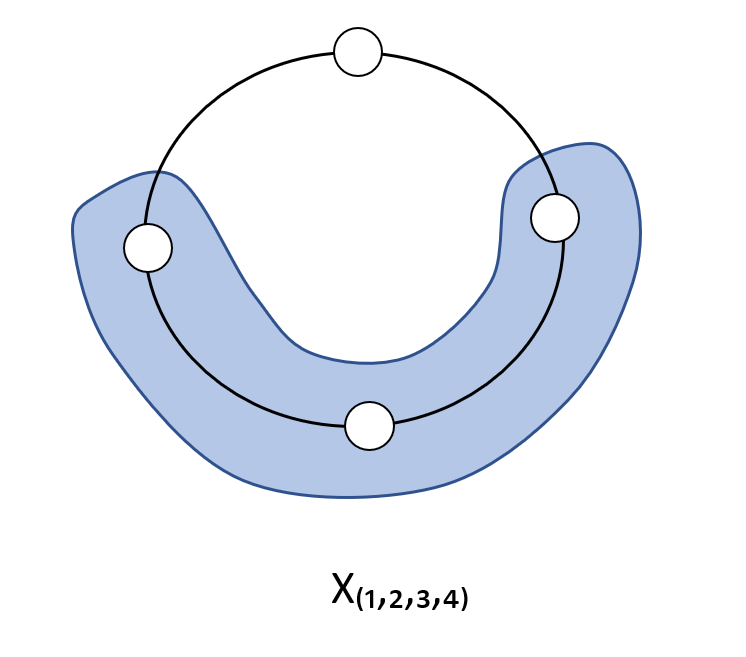}
	\caption{Example of tubing in $n=4$ case corresponding to $X_{(1,2,3,4)}$}
\end{figure}

\section{Polyhedral fans and \textit{g}-vector fans} \label{sec4}
In this section, we provide a brief introduction to polyhedral fans and \textit{g}-vector fans. One can then construct all possible polytopal realizations of any graph associahedra using these \textit{g}-vectors. We will use this method to construct Halohedron. It's pretty interesting to note that polytopal realizations of Halohedron based on the physics-inspired arguments proposed in \cite{salvatori20181loop} exactly match with this \textit{g}-vector fans based construction.

\subsection{Polyhedral fans}

A \textit{polyhedral fan} $\mathcal{F}$ is a collection of closed linear halfspaces. These linear halfspaces are called polyhedral cones $C$ and can be denoted as: 
\begin{equation}
C := \left\{ \sum_{\mathbf{r} \in \mathbf{R}}\lambda_r {\mathbf{r}} ~ | ~\lambda_r \in \R_{\geq 0} \right\}
\end{equation}
where, $\mathbf{R}$ is a set of vectors in $\R^n$ and $C$ represents a positive span of finitely many vectors of $\R^n$.  A co-dimension one face of cone $C$ is called ray. A cone is \textit{simplicial} if it is generated by a set of linearly independent vectors. The faces of polyhedral fan are the faces of its constituent cones.
A fan is simplicial if all of its cones are simplicial. It is complete if the union of its cones covers the entire space  $\R^n$.
\\

A \textit{polytope} $\mathbf{P}$ is a subset of $\R^n$ which can be defined in two equivalent ways:
\begin{itemize}
	\item As the intersection of finitely many closed linear halfspaces.
	\item Convex hull of finitely many points in $\R^n$.
\end{itemize}
The polytope $\mathbf{P}$ is simple if there are $dim(\mathbf{P})$ facets incident at each vertex. Consider a face F of polytope $\mathbf{P}$, the \textit{normal cone} of the face F is the cone generated by the rays normal to the facets of $\mathbf{P}$ which contains F. A \textit{normal fan} of $\mathbf{P}$ is a collection of \textit{normal cone}s of all of its faces. In Figure (7) the rays are the normal vectors to the facets of the polytope.
\begin{figure}[hbt!] 
	\centering
	\includegraphics[scale=0.5]{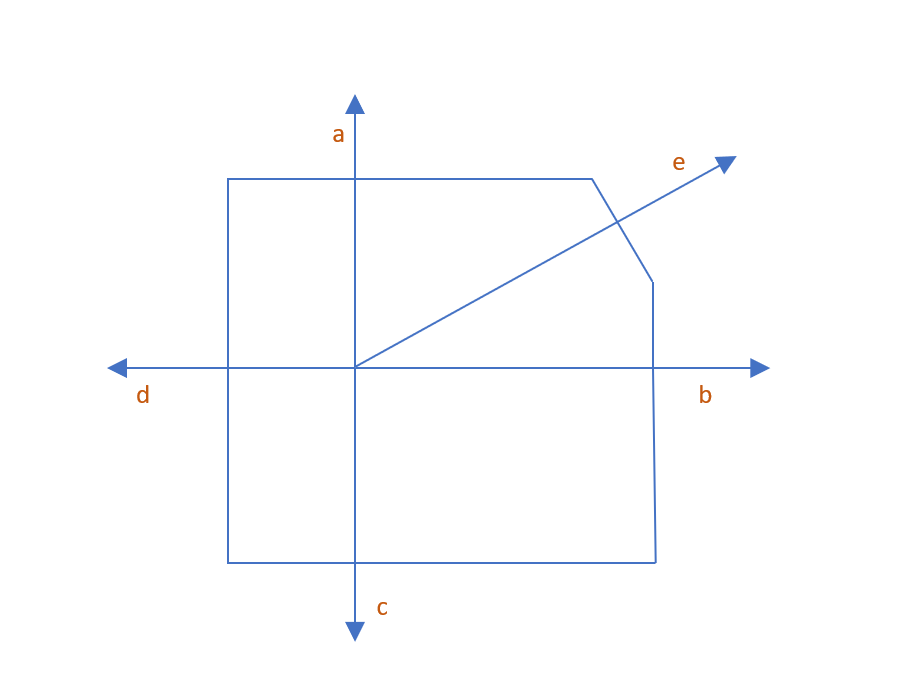}
	\caption{Normal fan with its rays \textit{a, b, c, d, e} and its polytopal realization} \label{fig:normalfan}
\end{figure}
\\

Now consider a simplicial, complete (union of all the cones is an entire space $\R^n$) and essential (contains null cone) fan $\mathcal{F}$.
Let $\textbf{\textit{G}}$ be a matrix whose rows are the N rays of $\mathcal{F}$, then for any height vector  \textbf{h} in $\R^N$, the fan $\mathcal{F}$  is the normal fan of the polytope defined as
\begin{equation}
\mathbf{P}_{\mathbf{h}}:=\{\textbf{x}\in \R^n|\textbf{\textit{G}x}\leq \textbf{h}\}
\end{equation}
and the rays of  fan are  normal vectors to  facets of the polytope.

\subsection{\textit{g}-vector fan}
In the last subsection, for some $\mathbf{h}\in \mathbb{R}^N$ , we obtained $\mathbf{P}_\mathbf{h}$ as the polytopal realization of normal fan such that the rays in the fan were normal vectors to the faces of the polytope. Next, our task would be to find a normal fan consisting of vectors corresponding to tubings of a graph and one can then expect that its polytopal realization will be graph associahedron.

Let \textbf{G} be a graph, denote the set of connected components as $\kappa(\mathbf{G})$ and the set of nodes as $V$. Using same notation as \cite{manneville2015compatibility}, we denote  $\mathcal{N}^{\Box}$ as the set of pairwise compatible design tubes also called \textbf{design nested complex}.
We define \textbf{g}-vectors of tubes as
\begin{equation}
\mathbf{g}(t):=\pi\left(\sum_{v\in t} e_v\right)
\end{equation} 
where $e_v$ are the basis vectors of $\R^V$ and $\pi$ is an orthogonal projection map onto the hyperplane $\mathbf{H}$ and the hyperplane is defined as 
\begin{equation}
\mathbf{H}:=\{\textbf{x}\in\mathbb{R}^V| \sum_{w \in W} x_w = 0 ~\textrm{ for all } W\in \kappa(\mathbf{G})\}
\end{equation}

For a design tube $t$ of $\mathbf{G}$, following \cite{manneville2015compatibility}, we set the \textit{g}-vectors 
\begin{equation}
\mathbf{g}(t):= \begin{cases} 
\sum_{v\in t} e_v \textrm{, t is a round tube}\\
-e_v, ~\textrm{ if t is a square tube }\\
\end{cases}
\end{equation} 
where, $e_v$ is the canonical basis of $\R^{V}$. 
Each basis vector $e_v$ maps to one of the vertex of the graph.
The collection of cones,
\begin{equation}
\mathcal{G}^{\Box}(\mathbf{G}):=\{\R_{\geq 0}(\mathbf{g}(T)) ~|~ T\textrm{ tubing on \textbf{G}}\}
\end{equation}
is a complete simplicial normal fan, called the design nested fan of $\mathbf{G}$, which realizes $\mathcal{N}^{\Box}$.
It has been proven in \cite{devadoss2010deformations}, that $\mathcal{G}^{\Box}(\mathbf{G})$ is polytopal and the polytope is obtained from the iterated truncation of cube called graph cubeahedron. For a height vector $\mathbf{h}\in\mathbb{R}^V$, the design nested fan $\mathcal{G}^{\Box}(\mathbf{G})$ is the normal fan of graph cubeahedron defined as, 
\begin{equation}\label{eq:polyre}
\mathcal{C}G:=\left\{ \mathbf{x}\in \mathbb{R}^V ~|~ \langle \mathbf{g}(t)|\mathbf{x}\rangle\leq \mathbf{h}_t~ \textrm{for any tube} ~t\in \mathcal{T}(G)\right\}
\end{equation}
where $\mathcal{T}(G)$ is the set of all possible tubings of the graph $\mathbf{G}$. The height vectors satisfy 
\begin{equation}
\mathbf{h}_{t \smallsetminus v}+\mathbf{h}_{t\smallsetminus{v'}}>~\mathbf{h}_t ~ + ~ \mathbf{h}_{t\smallsetminus\{v,v'\}}\label{eqn3:cons}
\end{equation} 
and $\mathbf{h}_{\emptyset}=\mathbf{h}_{\mathbf{G}}=0$. Height vector $\mathbf{h}_{t\smallsetminus v}$ corresponds to the facet associated to tube $t\smallsetminus v$, where $v,~v'$ are two non disconnecting nodes of the tube $t$. In\cite{manneville2015compatibility} three possible propositions for height functions satisfying (\ref{eqn3:cons}) are mentioned.

These height vectors form a cone called \textit{type-cone} and if \textit{type-cone} is simplicial it provides all possible realizations of  fan $\mathcal{F}$. The type cone is defined as
\begin{equation}
\mathbb{TC}(\mathcal{F}):=\left\{\mathbf{h}\in \mathbb{R}^N | \sum_{s \in \mathbf{R}\cup \mathbf{R}'}\alpha_{\mathbf{R},\mathbf{R}'}~\mathbf{h}_s>0\right\}
\end{equation}
where the pair of cones $\left\{ \mathbf{R}_{\geq0},\mathbf{R}'_{\geq0} \right\}$ forms an extremal adjacent pair and $\alpha$'s are constants. In figure (\ref{fig:normalfan}), $\{a,b\}$,$\{b,c\}$ and $\{c,d\}$ are extremal adjacent pairs. If there are $N-n$ such pairs, where $N$ is the dimension of the type-cone($i.e.$ there are $N$ rays of the normal fan) and $n$ is the dimension of the ambient space in which normal fan is defined, then the type-cone is simplicial and provides all polytopal realizations of the normal fan.

Indeed, type-cone of the normal fan $\mathbb{TC(\mathcal{F})}$ of Associahedron is simplicial and the polytopal realizations of Associahedron in \cite{Arkani_Hamed_2018} are the simplicial type cone based realization. But in general, type-cone of a normal fan of a graph associahedron is not simplicial so the type cone cannot provide all possible polytopal realizations. So rather in the construction of Halohedron, we use the $g$-vector fans approach.

\section{Construction of Halohedron} \label{sect5}
Finally, in this section, we arrive at the \textit{g}-vector fan-based polytopal realizations of Halohedron $H_n$ which are realized as graph cubeahedron for a cycle-graph with $n$ nodes. Below, we first give 2 examples for the cases of a graph with 2 nodes and 3 nodes respectively. Then we provide a polytopal realization of graph cubeahedron for cycle graph with  $n$ nodes. In this process, we find the height vectors of Salvatori's \cite{salvatori20181loop} construction.

\subsection{Case 1: Cycle-graph with 2 nodes}
\begin{figure}
	\centering
	\includegraphics[scale=.7,width=\linewidth]{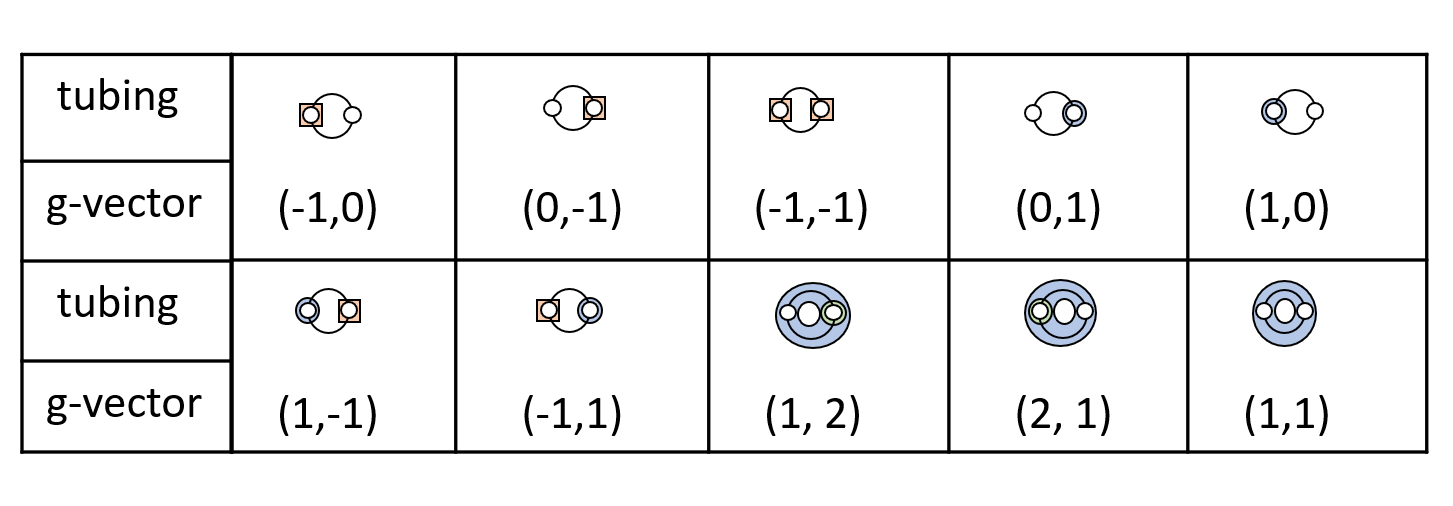}
	\caption{Table showing the \textit{g}-vectors for tubings of cycle-graph with 2 nodes}
\end{figure}
\begin{figure}[!hbt] 
	\centering
	\includegraphics[scale=.7]{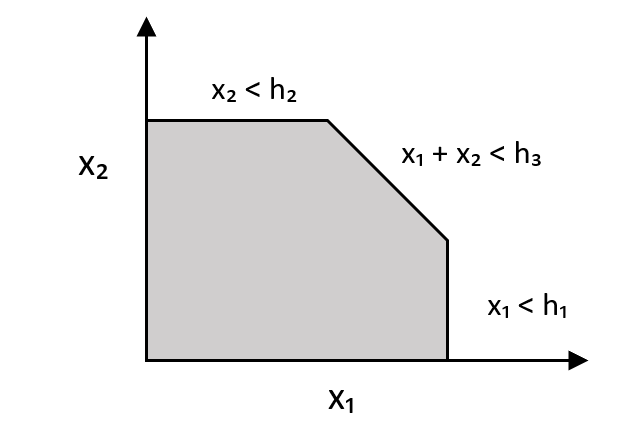}
	\caption{Facet defining inequalities for $H_2$}\label{fig:h2}
\end{figure}

The table in figure (8) shows \textit{g}-vectors corresponding to the tubings for a cycle-graph with 2 nodes. Substituting these \textit{g-vectors} in (\ref{eq:polyre}) we get the following facet defining inequalities of $H_2$:
\begin{equation} 
-x_1~\leq ~h_{-1} ;\quad -x_2~\leq~h_{-2}  ;\quad x_1~\leq~ h_{1} ;\quad  x_2~\leq~h_{2} ;\quad x_1+x_2~\leq h_{0}\label{eq2:n2}
\end{equation}
The corresponding Halohedron $H_2$ is shown is figure (\ref{fig:h2}).

For $n=2$, in Salvatori's construction $X_1$ and $X_2$ are the independent variables which are associated to the single square tube across the two nodes of the graph and the variables corresponding to the other facet defining inequalities of Halohedron are:
\begin{eqnarray}
X_{(1,2)}:=\epsilon^1_{(1,2)}-X_1\\
X_{(2,1)}:=\epsilon^1_{(2,1)}-X_2\\
X_0:=X_{(1,2)}+X_{(2,1)}-\epsilon_0^2
\end{eqnarray}
These lead to a convex polytope if all the variables 
are positive i.e. 
\begin{eqnarray} \label{eqn:n_2} 
X_1 \geq 0,   \quad  X_2 \geq 0\\
\epsilon^1_{(1,2)}-X_1\geq 0 \\
\epsilon^1_{(2,1)}-X_2\geq 0 \\
X_0:=X_{(1,2)}+X_{(2,1)}-\epsilon_0^2 \geq 0\label{eqn:n_21}
\end{eqnarray}

Comparing both the sets of variables in (\ref{eq2:n2}) and (\ref{eqn:n_2})-(\ref{eqn:n_21}), we note that  $h_{-1}=h_{-2}=0$, $h_1=\epsilon^1_{(1,2)},~ h_2=\epsilon^1_{(2,1)}$ and $h_0=\epsilon^1_{(1,2)}+\epsilon^1_{(2,1)}-\epsilon_0^2$. We will show in the next section that these height functions in terms of these $\epsilon$'s indeed satisfy (\ref{eqn3:cons}) $i.e.$
\begin{equation}\label{eqn:h2}
h_0~<~h_1+~h_2
\end{equation}
where $h_1$,$~h_2$ are the height functions corresponding to the tubes obtained by removing nodes $1$ , $2$ respectively from the round tube encircling both nodes.

Substituting $h$'s in terms of $\epsilon$'s in (\ref{eqn:h2}) we get,
\begin{eqnarray}
\epsilon^1_{(1,2)}+\epsilon^1_{(2,1)}-\epsilon_0^2 < \epsilon^1_{(1,2)}+\epsilon^2_{(2,1)}\\
\implies -\epsilon_0^2 <0
\end{eqnarray}
We will show in next section that $\epsilon_0^2$ satisfy the above constraint.

\subsection{Case 2: Cycle-graph with 3 nodes}

\begin{figure}
	\centering
	\includegraphics[scale=.7,width=\linewidth]{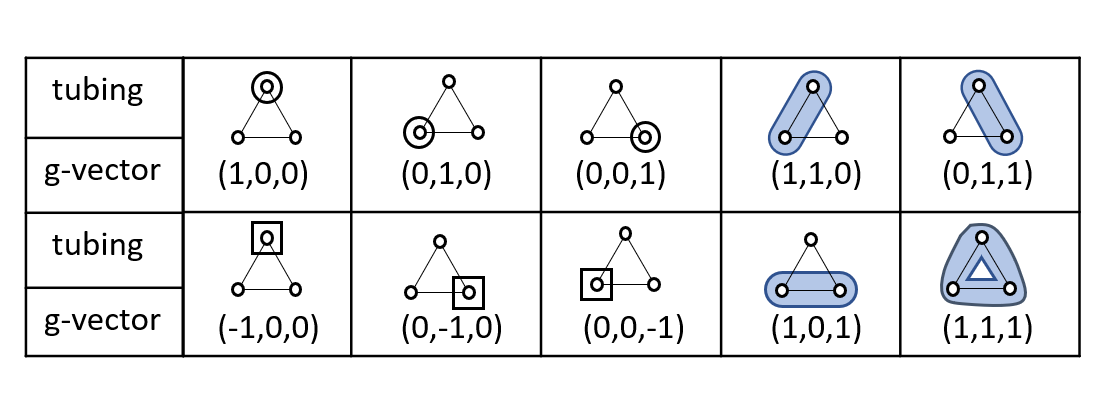}
	\caption{Table showing the \textit{g}-vectors for tubings of cycle-graph with 3 nodes}\label{fig:n3}
\end{figure}
Using the $g-vectors$ ( from figure (\ref{fig:n3})) for the cycle graph with 3 nodes in (\ref{eq:polyre}) we get the following inequalities:
\begin{eqnarray}\label{eqn2:n_3}
x_1+x_2+x_3\leq h_0; \quad x_1+x_2 \leq h_{(1,2)}; \quad x_2+x_3 \leq h_{(2,3)}; \quad x_3+x_1\leq h_{(3,1)}; \\ 
x_1\leq h_1; \quad x_2\leq h_2; \quad x_3\leq h_3; \quad -x_1 \leq h_{-1}; \quad -x_2 \leq h_{-2}; \quad -x_3 \leq h_{-3};\label{eqn2:n_31}
\end{eqnarray}
These inequalities carve out halohedron $H_3$ in $\R^3$. 

In Salvatori's construction $X_1$, $X_2$ and $X_3$ are the independent variables and the variables corresponding to the other facet defining linear functions are:
\begin{eqnarray}\label{eqn1:n_3}
X_{1,2}:=\epsilon^1_{1,2}-X_1 \quad
X_{2,3}:=\epsilon^1_{2,3}-X_2 \quad
X_{3,1}:=\epsilon^1_{3,1}-X_3 \\
X_{(1,2,3)}:=X_{1,2}+X_{2,3}-\epsilon^2_{(1,2,3)};\quad
X_{(2,3,1)}:=X_{2,3}+X_{3,1}-\epsilon^2_{(2,3,1)};\\
X_{(3,1,2)}:=X_{3,1}+X_{1,2}-\epsilon^2_{(3,1,2)}; \quad
X_0:=X_{1,2}+X_{2,3}+X_{3,1}-\epsilon_0^3\label{eqn1:n_31}
\end{eqnarray}
The region inside the Halohedron is defined by imposing positivity condition on all of the above variables.
After imposing positivity on all the variables in (\ref{eqn1:n_3})-(\ref{eqn1:n_31}) and comparing with the inequalities (\ref{eqn2:n_3})-(\ref{eqn2:n_31}), we obtain  relationship between the height functions and the $\epsilon$'s:
\begin{eqnarray}
h_{(1)}=\epsilon^1_{(1,2)}; \quad h_{(2)} =\epsilon^1_{(2,3)}; \quad h_{(3)} = \epsilon^1_{(3,1)};\\
h_{(1,2)}=\epsilon^1_{(1,2)}+\epsilon^1_{(2,3)}-\epsilon^2_{(1,2,3)};\quad h_{(2,3)}=\epsilon^1_{(2,3)}+\epsilon^1_{(3,1)}-\epsilon^2_{(2,3,1)};\\ h_{(3,1)}=\epsilon^1_{(3,1)}+\epsilon^1_{(1,2)}-\epsilon^2_{(3,1,2)};\quad
h_0=\epsilon^1_{(1,2)}+\epsilon^1_{(2,3)}+\epsilon^1_{(3,1)}-\epsilon^3_0
\end{eqnarray}

Again we have to prove that these height functions satisfy (\ref{eqn3:cons}) $i.e.$:
\begin{eqnarray}
h_0+h_{(2)}<h_{(1,2)}+h_{(2,3)}
\end{eqnarray}
and similarly other set of inequalities. 
Substituting $h$'s in terms of $\epsilon$'s we get:
\begin{eqnarray}
-\epsilon^3_0<-\epsilon^2_{(1,2,3)}-\epsilon^2_{(2,3,1)}
\end{eqnarray}
We will prove this in next section using the convexity property of the polytope.

\subsection{Case 3: Cycle-graph with n nodes}
We know that Halohedron consists of 3 kinds of facets: 
\begin{itemize}
	\item One cyclohedral facet $W_n$ which corresponds to the round tube enclosing the entire $cycle$-graph.
	\item $n^2-n$ copies of $K_m\times H_{n-m+1}$ which are associated to round tubes enclosing a set of consecutive nodes of graph.
	\item $n$ copies of $K_{n+1}$  which correspond to tubings consisting of only a square tube.  
\end{itemize}
The Halohedron is embedded in ${\R}^n$ with $(x_1,x_2\hdots,x_n)$ being any generic point in ${\R}^n$.
The cyclohedron facet is formed by the \textit{g}-vector 
\begin{align}
g &= \begin{pmatrix}
1\\
1 \\
\vdots \\
1
\end{pmatrix}
\end{align}
corresponding to the tube encircling the whole graph.    
This \textit{g}-vector gives rise to the facet inequality:
\begin{eqnarray}
\langle (1,1\hdots,1)|(x_1,x_2\hdots,x_3)\rangle \leq h_0\\
\implies x_1 + x_2 + \hdots +x_n \leq h_0
\end{eqnarray}
In Salvatori's construction this facet corresponds to the variable $X_0$ defined as:
\begin{equation}
X_0:=\sum_{i=1}^n X_{(i,i+1)}-\epsilon^n_0
\end{equation}
Using the definitions of $X_{(i,i+1)}$ from section 4 and the positivity of $X_0$ for all the points inside the Halohedron we obtain
\begin{equation}
\epsilon^1_{(1,2)}+\epsilon^1_{(2,3)}+\hdots+\epsilon^1_{(n,1)}-\epsilon^n_0>X_1+X_2+\hdots+X_n
\end{equation}
So we find $h_0$ in terms of $\epsilon$:
\begin{equation}
h_0=\epsilon^1_{(1,2)}+\epsilon^1_{(2,3)}+\hdots+\epsilon^1_{(n,1)}-\epsilon^n_0
\end{equation}

The \textit{g}-vector corresponding to factorisation facets $K_m\times H_{n-m+1}$ are of the form:
\begin{align}
g &= \begin{pmatrix}
0\\            
\vdots \\
1\\
1\\
\vdots\\
1\\ \vdots \\
0
\end{pmatrix}
\end{align}
There are $m$ consecutive ones in the column vector denoting the tube encircling m consecutive nodes of the cycle graph.
There will be $n^2-n$ such $g-vetors$. Using (\ref{eq:polyre}) we obtain the facet defining inequalities as :
\begin{eqnarray}\label{eqn2:n}
\langle (0,0,\hdots ,1,\hdots ,1,0\hdots ,0)|(x_1,x_2\hdots,x_n)\rangle \leq h_{(i,\hdots ,i+m)}\\
\implies x_i + x_{i+1} + \hdots +x_{i+m} \leq h_{(i,\hdots ,i+m)}
\end{eqnarray}
These correspond to the variables $X_I$ which are defined as:
\begin{equation}
X_I:=\sum_{j\in I'}X_{(j,j+1)}-\epsilon_I^{|I'|}
\end{equation}
where, $I$ denotes $ m+1$ consecutive nodes and $I'$ is the same set as $I$ with the last node removed.  Using the positivity condition on the variables and the definition of $X_{(i,i+1)}$
\begin{equation}\label{eqn1:n}
\sum_{j\in I'}\epsilon_{j,j+1}-\epsilon_I^{|I'|}>\sum_{j\in I'}X_j
\end{equation}

Comparing (\ref{eqn1:n}) with (\ref{eqn2:n}) we obatin the following relationship between the height functions $h$'s and $\epsilon$'s :
\begin{equation}
h_{(i,\hdots,i+m)}=\sum_{j\in I'}\epsilon_{j,j+1}-\epsilon_I^{|I'|}
\end{equation}
where $I'=\{i,\hdots,i+m\}$

Now we are left with $n$ associahedral facets which label the single square tubes across different nodes of the graph.
The corresponding $g-vectors$ are of the form:
\begin{align}
g &= \begin{pmatrix}
0\\            
\vdots \\
-1\\
\vdots\\
0
\end{pmatrix}
\end{align}
Again using (\ref{eq:polyre}) we obtain:
\begin{eqnarray}
\langle (0,\hdots 0,-1,0\hdots ,0)|(x_1,x_2\hdots,x_n)\rangle \leq h_{(-i)}\\
\implies -x_i \leq h_{(-i)}
\end{eqnarray}
In Salavatori's construction all these $h_{(-i)}$'s are zero. 

\section{Proof of the constraints on $\epsilon$'s}\label{sec6}

In the last section, we established equivalence between Salvatori's construction of Halohedron and the \textit{g}-vector fans based realization of Halohedron. As a result, we ended up with constraints on $\epsilon$'s which we didn't prove then. In this section, we will show that indeed these constants appearing in the linear facet defining functions do satisfy the condition (\ref{eq:polyre}) for height vectors. 

In the proof, we use the idea of adjacent facets of Halohedron $H_n$.
An intuitive way to infer which two facets of the Halohedron are adjacent is to have a picture of both the facets being associated with tubings (tubings are set of compatible tubes) and that tubings consisting of only round tubes are involved in the truncation of $n$-cube. If $v$ and $v'$ are two non-disconnecting nodes of the tube $t$, then $t \cup t\smallsetminus \{v,v'\}$  is a tubing but $t\smallsetminus \{v\} \cup t \smallsetminus \{v'\}$ is not a tubing as  $t\smallsetminus \{v\}$ intersects $t\smallsetminus \{v'\}$ so these are not compatible. 
Now since only tubings with round tubes are involved in truncation of $n$-cube thus forming the facets of Halohedra,  $t \cup t\smallsetminus \{v,v'\}$ does label a facet of Halohedron whereas $t    \smallsetminus \{v\} \cup ~t\smallsetminus \{v'\}$ does not label any facet. 
\begin{figure}[!hbt] \label{fig:proof}
	\centering
	\includegraphics[scale=.4]{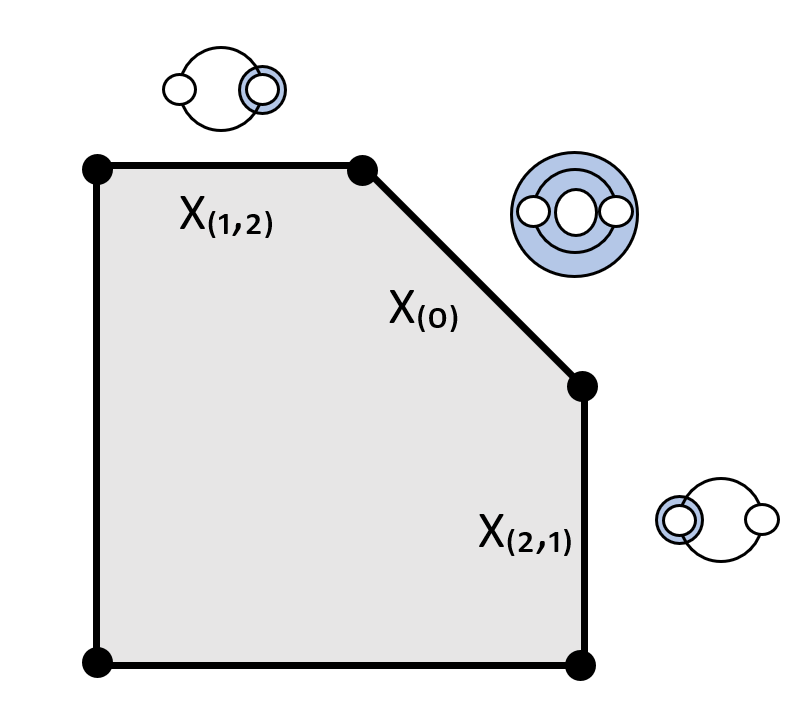}
	\caption{Illustrative example for $H_2$ showing $X_{1,2}$ and $X_{2,1}$ as non-adjacent facets.}\label{fig:proof}
\end{figure}
From the previous section we use the direct equivalence between the facet labelled by the \textit{g}-vector of the tubing and the variables $X_I$. Now let $X_I$, $X_J$, $X_K$, $X_L$ be the facets equivalent to those obtained from tubings $t$, $t    \smallsetminus \{v\}, ~t\smallsetminus \{v'\}$, $t\smallsetminus \{v,v'\}$ respectively. Following the above argument, the facet associated to the tubing $t \cup t\smallsetminus \{v,v'\}$  corresponds to the intersection of $X_I$ and $X_L$. So $X_I+X_L \geq 0$ is saturated on the Halohedron when both $X_I$ and $X_L$ simultaneously vanish whereas same is not true for $X_J+X_K$.

In figure (\ref{fig:proof}) edge $X_0$ corresponds to tube $t$; $X_{(1,2)}$ and $X_{(2,1)}$ corresponding to tubes $t\smallsetminus\{v\}$ and $t\smallsetminus\{v'\}$ are non-adjacent and intersect outside the Halohedron. So
\begin{equation}
X_{(1,2)}+X_{(2,1)}>0
\end{equation}

\subsection{n=2 case}
\textbf{To prove},
\begin{equation}
-\epsilon^2_0<0
\end{equation}
\textbf{Proof}:
The variables in $n=2$ case are $X_1,~X_2,~X_{(1,2)},~X_{(2,1)}$~and$ ~ X_0$ defined as 
\begin{eqnarray*}
	X_{(1,2)}:=\epsilon^1_{(1,2)}-X_1\\
	X_{(2,1)}:=\epsilon^1_{(2,1)}-X_2\\
	X_0:=X_{(1,2)}+X_{(2,1)}-\epsilon_0^2
\end{eqnarray*}

Now,
\begin{equation}
X_{(1,2)}+X_{(2,1)}\geq 0
\end{equation}
as both the variables are positive inside the polytope.
But $X_{(1,2)}$ and $X_{(2,1)}$ are the non-adjacent sides of the convex polytope, so these two facets cannot intersect at any point inside the polytope 
\begin{eqnarray}
\implies X_{(1,2)}+X_{(2,1)}>0 \\
X_{(1,2)}+X_{(2,1)}-\epsilon_0^2+\epsilon_0^2>0\\
\implies X_0>-\epsilon_0^2
\end{eqnarray}
and as $X_0\geq 0$ . Hence,
\begin{equation}
-\epsilon_0^2<0
\end{equation}

\subsection{n=3 case}
\textbf{To prove},
\begin{eqnarray}
-\epsilon^3_0<-\epsilon^2_{(1,2,3)}-\epsilon^2_{(2,3,1)}
\end{eqnarray}
\textbf{Proof}: The variables in $n=3$ case are $X_1,~X_2,~X_{(1,2)},~X_{(2,3)},~X_{(3,1)},~ X_{(1,2,3)},~X_{(2,3,1)},~X_{(3,1,2)},X_0$. They are defined as

\begin{eqnarray*}
	X_{1,2}:=\epsilon^1_{1,2}-X_1 &
	X_{2,3}:=\epsilon^1_{2,3}-X_2 &
	X_{3,1}:=\epsilon^1_{3,1}-X_3 \\
	X_{(1,2,3)}:=X_{1,2}+X_{2,3}-\epsilon^2_{(1,2,3)};&
	X_{(2,3,1)}:=X_{2,3}+X_{3,1}-\epsilon^2_{(2,3,1)};&
	X_{(3,1,2)}:=X_{3,1}+X_{1,2}-\epsilon^2_{(3,1,2)}\\
	&X_0:=X_{1,2}+X_{2,3}+X_{3,1}-\epsilon_0^3
\end{eqnarray*}

The non adjacent facets $X_{(1,2,3)}$ and $X_{(2,3,1)}$ intersect outside the convex polytope so these cannot be simultaneously zero for the points in  Halohedron, thus
\begin{equation}
X_{(1,2,3)}+X_{(2,3,1)}>0
\end{equation}
Substituting the variables above in terms of the independent variables $X_1,~X_2,~X_3$ we obtain 
\begin{eqnarray}
\epsilon^1_{(1,2)}-X_1 +\epsilon^1_{(2,3)}-X_2 +\epsilon^1_{(3,1)}-X_3+\epsilon^1_{(2,3)}-X_2-\epsilon^2_{(1,2,3)}-\epsilon^2_{(2,3,1)}>0\\\label{eqn3:n_3}
\implies X_0+X_{(2,3)}-\epsilon^2_{(1,2,3)}-\epsilon^2_{(2,3,1)}+\epsilon^3_0>0
\end{eqnarray}
Again we exploit the fact that $X_0$ and $X_{(2,3)}$ are the adjacent facets on the polytope so these can saturate the inequality
\begin{equation}
X_0+X_{(2,3)}\geq 0
\end{equation}
Using this relation in (\ref{eqn3:n_3}) we obtain
\begin{equation}
\epsilon^2_{(1,2,3)}+\epsilon^2_{(2,3,1)}-\epsilon^3_0<0
\end{equation}
which completes our proof.
One can similarly prove the inequalities for other sets of height functions.

\subsection{Abritrary n case}
Following (\ref{eqn3:cons}) for arbitrary $n-$case, the height vectors satisfy:
\begin{equation}\label{eqn:last}
h_{t\smallsetminus v} +h_{t\smallsetminus v'}>h_t +h_{t\smallsetminus \{v,v'\}}
\end{equation} 

Let $X_I$, $X_J$, $X_K$ and  $X_L$ be the variables which correspond to the facets which are in $1-1$ correspondence to the facets labelled by tubings $t$, ~$t\smallsetminus v$,~ $t\smallsetminus v'$ and $t\smallsetminus \{v,v'\}$ then substituting the height functions of Salvatori's construction in (\ref{eqn:last}) we obtain:
\begin{equation}
\epsilon_J^{|J'|}+\epsilon_K^{|K'|}-\epsilon_I^{|I'|}-\epsilon_L^{|L'|}<0
\end{equation}
which can be proved in the similar approach as for $n=2$ and $n=3$ case.



\section*{Acknowledgements}
I am very grateful to Alok Laddha for suggesting the problem; constant guidance, insightful discussions and various valuable comments on improving the manuscript. I would also like to thank Prashanth Raman for clarifying many points during several discussions that we had in the early stages of the project.


\end{document}